\documentclass[aps,preprint,tightenlines]{revtex4}
\usepackage{graphicx}
\newcommand{\sh} {/ \hskip-5pt }
\begin{document}
\preprint{\rm FIU-NUPAR-\today{}\\}

\title{Large  $Q^2$ Electrodisintegration of the Deuteron in  Virtual Nucleon Approximation}
\author{Misak M. Sargsian}
\affiliation{Florida International University, Miami, Florida 33199, USA\\ \\ \\ }

\begin{abstract}
The two-body break up of the deuteron is studied at high $Q^2$ kinematics,
with main motivation to probe the deuteron at small internucleon distances. 
Such studies are associated with the probing of high momentum component of the deuteron 
wave function. For this, two main theoretical issues have been 
addressed such as electromagnetic interaction of the virtual photon with the bound nucleon 
and the strong interaction of produced baryons in the final state of the break-up reaction. 
Within virtual nucleon approximation we developed a new prescription to account for 
the  bound nucleon effects in electromagnetic interaction.
The  final state interaction at high $Q^2$ kinematics is calculated within generalized  
eikonal approximation~(GEA).  
We studied the uncertainties involved in the calculation and performed comparisons 
with the first experimental data on deuteron electrodisintegration at large $Q^2$.
We demonstrate that  the experimental data confirm GEA's early prediction that the 
rescattering is maximal at $\sim 70^0$ of recoil nucleon production  relative to 
the momentum of the virtual photon.  Comparisons also show that the forward recoil nucleon 
angles are best suited for studies of the electromagnetic interaction of bound nucleons and 
the high momentum structure of the deuteron. Backward recoil angle kinematics show sizable 
effects due  to the $\Delta$-isobar contribution. The latter indicates the  importance of 
further development of GEA to account for the inelastic transitions in the intermediate 
state of the electrodisintegration reactions.
\end{abstract}
\maketitle

\section{Introduction}
Two-body electro-disintegration  of the deuteron at high $Q^2$  
represents a powerful tool for studying one of the most fundamental issues of 
nuclear physics such as nuclear forces at intermediate to short distances.
Despite all the successes in constructing interaction potentials for $NN$ scattering, 
the most advanced potentials\cite{NJMN,Argonne,Bonncd,Paris} still use phenomenological 
form-factors  to account for intermediate to short range  interactions.
Such form-factors shed little light on how   nuclear forces at short distances  follow 
from the basic concepts of QCD.
Presently only the long range $NN$ interaction  is understood on fundamental QCD grounds.

The situation is not spectacular also from the experimental point of view. New experiments aimed at 
studies of $NN$ interaction at short distances practically stopped after the 
reassignment of AGS at Brookhaven National Laboratory\footnote{The best hopes for 
the restart of the program of  high energy baryonic experiments are associated   
with the  upcoming projects of J-PARC, Japan\cite{JPARC} and GSI, Germany\cite{PANDA}.}.

In this respect an  alternative way of  studying nuclear forces at short distances is to probe 
$NN$ systems in nuclei at short space-time separations.
Expectations that this can be achieved only at high-momentum transfer  
reactions (see e.g.\cite{FS81,FS88,hnm,MS01,revsrc}) was confirmed in a series of 
experiments with high energy electron\cite{Day,Kim1,Kim2,Eip5,Eip6} and proton 
probes\cite{Yael,Eip1,Eip2,Eip3,Eip4}.
Some of the unique  results of these experiments were the observations of the  
scaling for the ratios of inclusive cross sections  of nuclei and the deuteron\cite{Day}  
(or $^3He$\cite{Kim1,Kim2}) at $x_{Bj}> 1$ at $Q^2> 1.5$~GeV$^2$ 
as well as the observation of the strong (by factor of 20) dominance of  
$pn$ relative to $pp$/$nn$ short-range correlations in the $^{12}C$ nucleus for bound nucleon momenta  
300-600~MeV/c\cite{Eip4,Eip5,Eip6}.  If the first result was an indication that two (or more) nucleons can be 
probed at small separations, the second one was an  indication of the dominance of the 
tensor-forces\cite{Tg2,Sch,ACM}  in such correlations.  

The simplest reaction which could be used to investigate short-range $NN$ interactions 
using nuclear targets is  the exclusive electrodisintegration of the deuteron in which 
large magnitudes of the  relative momentum of the  $pn$ system  in the ground state are  probed.   
Three experiments\cite{Ulmer,Kim3,Werner} have already been performed using the relatively 
high~(up to 6~GeV) energy electron beam of the Jefferson Lab and more comprehensive experimental 
program will follow after the 11~GeV upgrade of the lab\cite{JLab12}.

The prospect of having  detailed experimental data on high energy deuteron electrodisintegration
makes the development of theoretical approaches for description of these reactions  a pressing issue.

One of the first models for  high energy break-up of the deuteron were developed 
in the mid 90's in which main emphasis was given to the studies of nucleon rescattering in the 
final state of the reactions\cite{pdppn,Sabine1,edepn95,pdppnct}.
These models were simple in a way that they assumed a factorization of the electromagnetic  
$\gamma^*N$ and final state $NN$ interactions and  
considered the rather small values of relative momenta of the bound $pn$ system. 

The extension of these calculations to a larger internal momentum region required more elaborate
approaches and several studies were done in this direction\cite{Sabine2,Ciofi1,MS01,Ciofi2,Sabine3,Sabine4,
SabineOff_shell,Laget}.

In this work we  develop a theoretical  model  for the description of high $Q^2$  exclusive electrodisintegration of 
the deuteron in knock-out kinematics  based on  virtual nucleon approximation.   The main theoretical 
framework is based on the generalized eikonal approximation~(GEA)\cite{edepn95,gea,MS01,Ciofi2,Tg1,Ciofi3,Ciofi4} 
which allows us to represent  the scattering amplitude in the covariant form using effective Feynman diagram 
rules. In this way all  the virtualities involved in the scattering amplitudes are defined unambiguously.  
Reducing these amplitudes by choosing positive energy projections for the nucleon propagators allow
us to represent them  through the convolutions of the deuteron wave function, on- and off- shell components of 
electromagnetic current and $pn$ rescattering amplitude.  In addition to accounting for the off-shell effects,  
nonfactorized approximation is applied  to the  electromagnetic and $NN$ rescattering parts in the calculation 
of the final state interaction~(FSI)  amplitude.   
As a result our calculation extends beyond the distorted wave impulse approximation limit.
We also estimated the charge exchange contribution in the final state interaction in addition to the 
$pn\rightarrow pn$ rescattering part of the FSI amplitude included in the eikonal approximation.

The paper is organized as follows.  In Sec.~2  we briefly discuss the kinematics of 
the disintegration reaction which we consider most efficient in probing  the $pn$ system at 
small separations. Then we discuss the basic assumptions of virtual  nuclear approximation and proceed 
with the derivation of scattering amplitudes and the differential cross section of the reaction.

In Sec.~3,  after deriving the total scattering amplitude we performed a detailed theoretical analysis to 
identify the extent of uncertainties due to the off-shell part of the final state interaction as well as 
contribution due to charge-exchange rescattering. We also analyzed the role of the off-shell effects in 
the electromagnetic current of the bound nucleon. 
These analyses allowed us to conclude that at sufficiently large values of $Q^2$ ($\sim 6$~GeV$^2$) the 
three most important contributions into the disintegration process are the off-shell electromagnetic 
current of the bound nucleon, the deuteron wave function and the  on-shell part of the $NN$ scattering amplitude.

Furthermore we compare our calculations with the first available  high $Q^2$ experimental data.
These comparisons allow us to confirm the early prediction of GEA that the maximal strength of FSI corresponds to 
$\sim 70^0$ production of the recoil nucleon relative to the direction of the virtual photon. 
We also found  that forward angles of recoil nucleon are best suited for studies of the 
off-shell electromagnetic current and the deuteron wave function.  
Another observation is that in backward direction there is a sizable 
contribution due to the $\Delta$-Isobar production at the intermediate state of the reaction.
In Sec.~4 we give conclusions and an outlook on further development of the model. 
 
 \section{Cross Section of the Reaction}
\subsection{Kinematics}
We discuss the process:
\begin{equation}
  \label{reaction}  
e + d \rightarrow e^\prime + p + n,
\end{equation}
in knock-out kinematics in which case one nucleon (for definiteness we consider it to be a proton) absorbs  
the virtual photon and carries almost all its  momentum. The optimal kinematics for 
probing the initial $pn$ system at close distances is defined as follows:
\begin{eqnarray}
  \mbox{(a)}  \ Q^2 \ge 1 \ \mbox{GeV}^2 ; \ \ \ \    \mbox{(b)} \  \vec p_f  \approx \vec q ; \ \ \ \  
  \mbox{(c)} \  \ p_f \gg p_r  \ge 300 \  \mbox{MeV/c},
\label{kinematics}
\end{eqnarray}
where we define $q \equiv (q_0,\vec q)$,  $p_f \equiv (E_f,\vec p_f)$  and $p_r = (E_r,\vec p_r)$ as four-momenta 
 of virtual photon, knock-out proton and recoil neutron respectively. Also $Q^2 = |\vec q|^2 - q_0^2$.
Conditions (\ref{kinematics})(b) and (c) define  the knock-out process, while  
condition (\ref{kinematics})(a) is necessary to satisfy Eq.(\ref{kinematics})(c).  
From the point of view of  the dynamics  of the reaction  one also needs Eq.(\ref{kinematics})(a)  
in order to provide 
a necessary resolution for probing the deuteron at small inter-nucleon distances.  

\begin{figure}[t]
  \centering\includegraphics[scale=1]{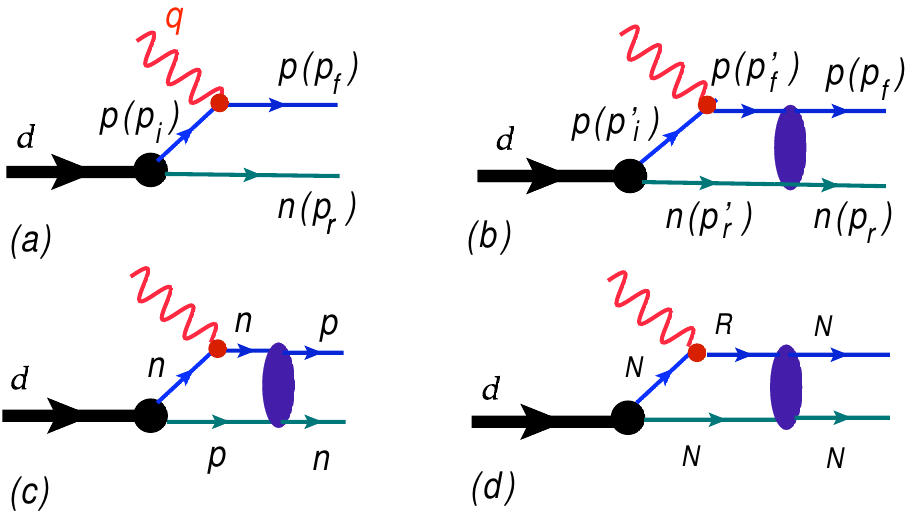}
  \caption{GEA diagrams}
\label{fig1}
\end{figure}

In the most simple picture the kinematics of Eq.(\ref{kinematics}) 
represent a scenario in which the high energy virtual photon removes the  proton from the $pn$ system 
leaving the neutron with the pre-existing relative momentum $p_r$ (see Fig.(\ref{fig1})a).
 
\subsection{Main Assumptions of Virtual Nucleon Approximation}

The virtual nucleon approximation is based on the following  main assumptions which also define
the limits of its validity. First,  one considers only the $pn$  component of the deuteron,  
neglecting inelastic initial state transitions.
Since the deuteron is in a isosinglet state this will correspond to restricting  the kinetic energy of 
recoil nucleon to 
\begin{equation}
T_N < 2(m_\Delta-m_N) \sim (m_{N^*}-m) \sim 500~\mbox{MeV}
\label{asum1}
\end{equation}
where $m$,  $m_\Delta$ and $m_{N^*}$ are masses of the nucleon and  low-lying  non-strange baryonic resonances, 
$\Delta(1232)$ and $N^*(1525)$. We neglect also the pionic degrees of freedom in the deuteron wave function. 
However we expect that the overall error introduced by this approximation to be small since the probability of 
low momentum pionic degrees of freedom is suppressed due to pseudogoldstone nature of pions in QCD as well 
as observation that $\pi NN$ form factors are soft (see e.g. discussion in Refs.\cite{FS88,revsrc}).

The second assumption is that the negative energy projection of the virtual nucleon propagator gives negligible 
contribution to  the scattering amplitude.  Such an assumption  can be justified if, 
\begin{equation}
M_d - \sqrt{m^2+ p^2}  > 0,
\label{asum2}
\end{equation}
where $M_d$ is the mass of the deuteron and $p$ is the relative momentum of the bound  $pn$ system.

The above two conditions can be satisfied if we restrict the momentum of the recoil neutron, $p_{r}\le 700$~MeV/c.
However due to the fact that we explicitly left out the non-nucleonic components of deuteron wave function, 
the momentum 
sum rule is not satisfied in virtual nucleon approximation (see discussions in Refs.(\cite{FS88,SSS02,polext}).

The third assumption which is made in the calculations  is that at  large   $Q^2$~($> 1$~GeV$^2$)  
the interaction of virtual photon with exchanged mesons are a small correction and 
can  be neglected (see e.g. discussions in Ref.\cite{MS01,hnm}).

\subsection{Generalized Eikonal Approximation}

The assumptions discussed above allow us to restrict the consideration  by the set of Feynman diagrams 
presented in Fig.(\ref{fig1}).  One can calculate these diagrams based on the effective Feynman diagram rules 
discussed in Ref.\cite{MS01}.  These rules allow us to formulate scattering amplitudes in the covariant form 
which unambiguously accounts for  all the  off-shell effects. Then 
we reduce the covariant amplitudes  into a non-covariant form  by choosing the positive 
energy projection of nucleon (or baryonic resonance)  propagators  at the intermediate state of the reaction.

Fig.\ref{fig1}(a) diagram corresponds to 
the plane wave  impulse approximation~(PWIA) in which the virtual photon knocks out one of the  nucleons 
from the deuteron leaving the second nucleon in  
the on-shell positive energy state.  Two nucleons do not interact in the final state of the reaction 
representing two outgoing plane waves.  The diagram of Fig.\ref{fig1}(b) represents a situation 
in which the  elastic electroproduction is followed by the elastic $pn \to pn$ rescattering.  
In this case the rescattering is forward in the sense that, for example, the proton struck by 
the virtual photon will 
attain its large momentum   after the  $pn\to pn$ rescattering. 
The amplitude of this scattering will be  referred to  as  a forward FSI amplitude.

The diagram of  Fig\ref{fig1}(c) corresponds to the scenario in which final state interaction proceeds through the 
charge-exchange $pn\to np$ rescattering.  In this case the final fast proton emerges from the process
in which the photon strikes the neutron which then  undergoes  a  $np\to pn$ charge-exchange 
rescattering. The amplitude of this scattering process  will be  referred to as a charge-exchange FSI amplitude.

The fourth diagram (Fig.\ref{fig1}(d)) corresponds to the electroproduction of 
an excited state $R$ with a subsequent $RN\to NN$ final state  rescattering.  The most important 
contribution to the fourth
diagram is due to  the $\Delta$-isobar~(IC), whose production threshold is closest to the quasielastic 
scattering kinematics.  Several factors make IC contribution small in high $Q^2$ limit at $x\ge 1$\cite{MS01,hnm}. 
One factor is the large longitudinal momenta of the initial nucleon involved in the $\Delta$-electroproduction  process:
\begin{equation}
p^{IC}_{i,z} = (1-x)m - {m^2_\Delta- m^2\over 2q},
\label{pzic}
\end{equation}
which indicates that choosing $x>1$ one can suppress the  electroproduction of $\Delta$-resonance in the 
intermediate states due to the large values of initial momenta entering in the deuteron  
wave function.

An additional suppression of $IC$ follows from the smallness of the $\gamma^* N\to \Delta$  transition form-factors 
as compared to the elastic form factors at $Q^2 \ge few$~GeV$^2$\cite{Stoler,Ungaro}.  Finally, due to the fact 
that the $\Delta N\to NN$ amplitude is dominated by pion or $\rho$-type reggeon exchanges,  it will be additionally 
suppressed by at least the factor of ${1\over \sqrt{Q^2}}$.   
In any case this contribution can be calculated in a selfconsistent 
way within the generalized eikonal approximation.  The calculation of these types of 
diagrams within GEA  will be  presented elsewhere\cite{MSinp}.

Below we will discuss the calculations of only the first three diagrams of Fig.\ref{fig1}.

\subsubsection{Plane Wave Impulse Approximation Amplitude}

The amplitude of the PWIA diagram in the covariant form can be written as follows: 
\begin{equation}
\langle s_f,s_r \mid A_0^\mu\mid s_d\rangle = -\bar u(p_r,s_r)\Gamma_{\gamma^* p}^\mu \frac{\sh p_i + m}{ p_i^2-m^2} 
\cdot \bar u(p_f,s_f) \Gamma_{DNN}\cdot \chi^{s_d},
\label{A0_1}
\end{equation}
where $\Gamma_{\gamma^* p}$ is the electromagnetic vertex of the $\gamma^*N\rightarrow N$ scattering and the vertex 
function  $\Gamma_{DNN}$ describes the transition of the deuteron into the $pn$ system. The notations $s_d, s_f, s_r$ 
describe the spin projections of the deuteron, knock-out proton and recoil neutron respectively.  The  spin function of the 
deuteron is represented by $\chi^{s_d}$. 
The four-momentum of the struck nucleon in the initial state within PWIA is defined as:
\begin{equation}
p_i = (E_d - E_r, \vec p_d - \vec p_r)   = (M_d - E_r, -\vec p_r)\mid_{LaB}.
\label{p_i_pwia}
\end{equation}
The above relation clearly shows the off-shell character of the struck nucleon in the initial state, 
since $p_i^2 \ne m^2$. Therefore expressing the initial nucleon's propagator through the on-mass shell nucleonic 
spinors is not valid.

However, using an approximation in which only positive energy projections are taken into account, one can 
isolate the on-shell part of the propagator by adding and subtracting  $E_i^{on}\gamma^0$ term to $\sh p_i$ 
as follows:
\begin{equation}
\sh p_i + m = \sh p_i^{on} + m  + (E_i^{off}-E_i^{on})\gamma^0,
\label{sumrule0}
\end{equation}
where $E_i^{off}= M_d-\sqrt{m_n^2+p_r^2}$ and $E_i^{on} = \sqrt{m_p+p_r^2}$ where $m_n$ and $m_p$ are the masses of 
the proton and neutron respectively\footnote{In further discussion we neglect the difference between proton and 
neutron  masses denoting them by $m$.}. Now we can separate the PWIA amplitude into on- and off-shell parts in the following way:
\begin{equation}
\langle s_f,s_r \mid A_0^\mu\mid s_d\rangle  = \langle s_f,s_r \mid A_{0,on}^\mu\mid s_d\rangle  + 
\langle s_f,s_r \mid A_{0_{off}}^\mu\mid s_d\rangle, 
\label{A0_2}
\end{equation}
where
\begin{equation}
\langle s_f,s_r \mid A_{0,on}^\mu\mid s_d\rangle  = -\bar u(p_r,s_r)\Gamma_{\gamma^* p}^\mu \frac{\sh p^{on}_i + m}{ p_i^2-m^2} 
\cdot \bar u(p_f,s_f) \Gamma_{DNN}\chi^s_d,
\label{A0_on}
\end{equation}
and 
\begin{equation}
\langle s_f,s_r \mid A_{0,off}^\mu\mid s_d\rangle  = -\bar u(p_r,s_r)\Gamma_{\gamma^* p}^\mu 
\frac{(E_i^{off}-E_i^{on})\gamma^0 }{ p_i^2-m^2} 
\cdot \bar u(p_f,s_f) \Gamma_{DNN}\chi^s_d.
\label{A0_off}
\end{equation}
For the on-shell part of the amplitude, using
\begin{equation}
\sh p_i^{on} + m = \sum\limits_{s_i} u(p_i,s_i)\bar u(p_i,s_i)
\label{sumrule}
\end{equation}
and the definition\cite{Gribov,Bertuchi},
\begin{equation}
\Psi_d^{s_d}(s_1,p_1,s_2,p_2) = - \frac{ \bar u(p_1,s_1)\bar u(p_2,s_2)\Gamma_{DNN}^{s_d} \chi_{s_d}} 
{(p_1^2 - m^2) \sqrt{2}\sqrt{(2\pi)^3 (p_2^2+m^2)^{1\over 2}}}
\label{wf}
\end{equation}
one obtains
 \begin{equation}
\langle s_f,s_r \mid A_{0,on}^\mu\mid s_d\rangle = \sqrt{2}\sqrt{(2\pi)^3 2 E_r}\sum\limits_{s_i} J_{N,on}^\mu(s_f,p_f;s_i,p_i)
\Psi_d^{s_d}(s_i,p_i,s_r,p_r),
\label{A0_on_1}
\end{equation}
where
\begin{equation}
J_{N,on}^\mu(s_f,p_f;s_i,p_i) = \bar u(p_f,s_f)\Gamma^\mu_{\gamma^*N} u(p_i,s_i).
\label{J_on}
\end{equation}
For the off-shell part  of the scattering amplitude one observes that the following relation,
\begin{equation}
\bar u(p_2,s_2)\Gamma_{DNN}\chi^{s_d} = - \sum\limits_{s_1}\frac {u(p_1,s_1)\Psi^{s_d}_d(s_1,p_1,s_2,p_2)}{2m}
(p_1^2-m^2)\sqrt{2}\sqrt{(2\pi)^3 2 (p_1^2+m^2)^{1\over 2}}
\label{off_shell_rel}
\end{equation}
satisfies Eq.(\ref{wf}).  Inserting it into Eq.(\ref{A0_off}) one obtains
\begin{equation}
\langle s_f,s_r \mid A_{0,off}^\mu\mid s_d\rangle  =  \sqrt{2}\sqrt{(2\pi)^3 2 E_r}\sum\limits_{s_i} J_{N,off}^\mu(s_f,p_f;s_i,p_i)
\Psi_d^{s_d}(s_i,p_i,s_r,p_r),
\label{A0_off1}
\end{equation}
where 
\begin{equation}
J_{N,off}^\mu(s_f,p_f;s_i,p_i) = \bar u(p_f,s_f)\Gamma^\mu_{\gamma^*N} \gamma^0 u(p_i,s_i) \frac {E_i^{off} - E_i^{on}}{2m},
\label{J_off}
\end{equation}
and $E_i^{off} = M_d - E_i^{on}$ and $E_i^{on} = \sqrt{m^2 + p_i^2}$.

One can combine on-and off-shell parts of the PWIA scattering amplitudes in the 
following form:
\begin{equation}
\langle s_f,s_r \mid A_{0}^\mu\mid s_d\rangle  =  \sqrt{2}\sqrt{(2\pi)^3 2 E_r}\sum\limits_{s_i} J_{N}^\mu(s_f,p_f;s_i,p_i)
\Psi_d^{s_d}(s_i,p_i,s_r,p_r),
\label{A0_3}
\end{equation}
where 
\begin{equation}
J_{N}^\mu(s_f,p_f;s_i,p_i) = J_{N,on}^\mu(s_f,p_f;s_i,p_i)+J_{N,off}^\mu(s_f,p_f;s_i,p_i).
\label{J_N}
\end{equation}
The above form of the electromagnetic current together with Eqs.(\ref{J_on}) and (\ref{J_off})  
represents our off-shell approximation. It is worth noting that 
the first, ``on-shell''  part of this current corresponds to the widely used ``De Forest'' 
approximation\cite{DeForest}. 
In the ``DeForest'' approximation
because of the absence of the second term  the gauge invariance is violated and 
the current conservation is restored by expressing $J_0$ or $J_{z}$ components through each other with different 
assumptions for the nucleon spinors and  electromagnetic vertices. 
The latter introduces more uncertainty since imposed relations are not unique. As a result one generates several 
forms of the off-shell electromagnetic currents\cite{DeForest}.

The additional ``off-shell'' part of the electromagnetic current in Eq.(\ref{J_N})  obtained in our approximation 
reduces the uncertainty of choosing on-shell nucleon spinors which is inherent to the ``De Forest'' approximation.  
The total current in Eq.(\ref{J_N}) is conserved since it is derived from the gauge invariant amplitude of
Eq.(\ref{A0_1}).
Therefore our approximation does not violate gauge invariance 
and no additional conditions  are needed to restore the  current conservation.

Note that our approximation is analogous to the one used in hadronic physics within light-cone 
approximation~(see e.g. \cite{BL80}) in which case an off-shell ``$\gamma^+$'' component of 
the fermion propagator is isolated in the similar manner as it is done for  the $\gamma^0$ component 
in our case.

\subsubsection{Forward Elastic Final State Interaction Amplitude}
We start by applying effective Feynman diagram rules to the diagram of Fig.(\ref{fig1})b, which yields:
\begin{eqnarray}
\langle s_f,s_r \mid A_{1}^\mu\mid s_d\rangle  & = &   -\int {d^4p_r^\prime \over i (2\pi)^4}  
\frac { \bar u(p_f,s_f)\bar u(p_r,s_r) F_{NN}[\sh p^\prime_r + m][\sh p_d - \sh p^\prime_r+\sh q+m]}
{(p_d-p^\prime_r+q)^2-m^2+i\epsilon}
\nonumber \\
& & \times \frac {\Gamma_{\gamma^*N}[\sh p_d-\sh p^\prime_r + m]\Gamma_{DNN}\chi^{s_d}}{((p_d-p^\prime_r)^2-m^2+i\epsilon)
(p^{\prime 2}_r - m^2 + i\epsilon)},
\label{A1_0}
\end{eqnarray}
were $F_{NN}$ represents the invariant $pn\rightarrow pn$ scattering amplitude that can be expressed as follows:
\begin{equation}
F_{NN}(s,t) = \sqrt{s(s-4m^2)}f_{NN} (s,t)
\end{equation}
where $s$ is the total invariant energy of the scattering $pn$ system and the $f_{NN}$ scattering amplitude 
is defined in such a way that  $Im f_{NN} = \sigma_{tot}$.
Furthermore we will use the following four-vectors defined as
\begin{equation}
p^\prime_i = p_d - p^\prime_r \ \ \mbox{and} \ \ p^\prime_f = p_d - p^\prime_r + q.
\end{equation}
We first integrate by $d^0 p^\prime_r$ through the positive energy pole of the spectator nucleon propagator at the 
intermediate state;
\begin{equation}
\int \frac {d^0 p^\prime_r}{p^{\prime2}_r-m^2+i\epsilon} = -i{2\pi\over 2E^\prime_r}.
\end{equation}
This integration allows us to use $\sh p^\prime_r+m =   \sum\limits_{s^\prime_r} u(p^\prime_r,s^\prime_r)\bar u(p^\prime_r,s^\prime_r)$. 
For $\sh p_d-\sh p^\prime $  we use a relation similar to Eq.(\ref{sumrule0}).  The same could be done for 
 $\sh p_d-\sh p^\prime_r +\sh q$. However for large values of $q$ the off-shell part in  Eq.(\ref{sumrule0}) 
is suppressed by ${|\vec q|-q_0\over |\vec q|}$ and in 
large $Q^2$ limit it's contribution is negligible. Thus we can use  
the on-shell relation, $\sh p_d-\sh p^\prime +\sh q = \sum\limits_{s^\prime_f} u(p^\prime_f,s^\prime_f)\bar u(p^\prime_f,s^\prime_f)$
for the knock-out nucleon spinors in the intermediate state. 
Using the above representations of the spinors,  the  definitions of the deuteron wave function~(Eq.(\ref{wf})) and 
electromagnetic current~(Eq.(\ref{J_N})) for the  scattering amplitude of Eq.(\ref{A1_0}) we obtain:
\begin{eqnarray}
& & \langle s_f,s_r \mid A_{1}^\mu\mid s_d\rangle  =     -\sqrt{2}(2\pi)^{3\over 2}\sum\limits_{s^\prime_f,s^\prime_r,s_i} 
\int {d^3p_r^\prime \over i (2\pi)^3}  \frac{\sqrt{2E^\prime_r}\sqrt{s(s-4m^2)}} {2E^\prime_r ((p_d-p^\prime_r+q)^2-m^2+i\epsilon)}  
\times \nonumber \\
& & \ \ \ \ \ \  \langle p_f,s_f;p_r,s_r\mid f^{NN}(t,s)\mid p^\prime_r,s^\prime_r;p^\prime_f,s^\prime_f\rangle \cdot  
J_{N}^\mu(s^\prime_f,p^\prime_f;s_i,p_i)
\cdot  \Psi_d^{s_d}(s_i,p^\prime_i,s^\prime_r,p^\prime_r).
\end{eqnarray}
Next, we consider the propagator of the knock-out proton in the intermediate state, using 
the condition of quasielastic scattering
\begin{equation}
(q + p_d - p_r)^2 = p_f^2 = m^2,
\end{equation}
one obtains
\begin{equation}
(p_d - p^\prime_r + q)^2 - m^2 + i\epsilon = 2|{\bf q}|(p^\prime_{r,z} - p_{r,z} + \Delta + i\epsilon),
\end{equation}
where
\begin{equation}
\Delta = {q_0\over |{\bf q}|}(E_r - E^\prime_r) + {M_d\over |{\bf q}|}(E_r - E^\prime_r)  + {p^{\prime 2}_r-m^2\over 2  |{\bf q}|}.
\label{Delta}
\end{equation}
Furthermore using the relation
\begin{equation}
{1\over (p^\prime_{r,z} - p_{r,z} + \Delta + i\epsilon} = -i \pi \delta(p^\prime_{r,z} - (p_{r,z} - \Delta)) + 
{\cal P}\int {1\over p^\prime_{r,z} - (p_{r,z}-\Delta) }
\end{equation}
and performing integration over $p^\prime_{r,z}$  we split the scattering amplitude into two terms; 
one containing  on-shell and the other off-shell  $pn\rightarrow pn$ scattering amplitudes as follows:
\begin{eqnarray}
& & \langle s_f,s_r \mid A_{1}^\mu\mid s_d\rangle  =     {i \sqrt{2}(2\pi)^{3\over 2}\over 4} \sum\limits_{s^\prime_f,s^\prime_r,s_i} 
\int {d^2p_r^\prime \over  (2\pi)^2}  \frac{\sqrt{2\tilde E^\prime_r}\sqrt{s(s-4m^2)}} {2\tilde E^\prime_r |q|}  \times \nonumber \\
& &  \ \ \ \ \ \  \langle p_f,s_f;p_r,s_r\mid f^{NN,on}(t,s)\mid \tilde p^\prime_r,s^\prime_r; \tilde p^\prime_f,s^\prime_f\rangle 
\cdot  J_{N}^\mu(s^\prime_f,p^\prime_f;s_i,\tilde p^\prime_i) \cdot  \Psi_d^{s_d}(s_i,\tilde p^\prime_i,s^\prime_r,\tilde p^\prime_r) 
\nonumber  \\
& & -  {\sqrt{2}(2\pi)^{3\over 2}\over 2} \sum\limits_{s^\prime_f,s^\prime_r,s_i} {\cal P}\int {dp^\prime_{r,z}\over 2\pi} 
\int {d^2p_r^\prime \over  (2\pi)^2}  \frac{\sqrt{2E^\prime_r}\sqrt{s(s-4m^2)}} {2E^\prime_r |{\bf q}|} 
\times \nonumber \\
 & & \ \ \ \ \ \  {\langle p_f,s_f;p_r,s_r\mid f^{NN,off}(t,s)\mid p^\prime_r,s^\prime_r;p^\prime_f,s^\prime_f\rangle
\over  p^\prime_{r,z}- \tilde p^\prime_{r,z} }  J_{N}^\mu(s^\prime_f,p^\prime_f;s_i,p^\prime_i)
\cdot  \Psi_d^{s_d}(s_i,p^\prime_i,s^\prime_r,p^\prime_r),
\label{a1_fsi}
\end{eqnarray}
where $\tilde p^\prime_r = (p_{r,z}-\Delta, p^\prime_{r,\perp})$, $\tilde E^\prime_r = \sqrt{m^2 + \tilde p^{\prime 2}_r}$, 
$\tilde p^\prime_i = p_d - \tilde p^\prime_r$ and $\tilde p^\prime_f = \tilde p^\prime_i + q$. 

For numerical estimates of the above amplitudes one needs on- and off-shell $pn\to pn$ amplitudes as an input. 
In high energy limit in which  the helicity conservation of small angle NN scattering is rather 
well established the  on-shell amplitude  is  predominantly imaginary and can be parameterized in the form
\begin{equation}
 \langle p_f,s_f;p_r,s_r\mid f^{NN,on}(t,s)\mid \tilde p^\prime_r,s^\prime_r; \tilde p^\prime_f,s^\prime_f\rangle  
= \sigma_{tot}^{pn}(i + \alpha) e^{{B\over 2}t} \delta_{s_f,s^\prime_f}\delta_{s_r,s^\prime_r},
\label{fnn_on}
\end{equation}
where $\sigma^{pn}_{tot}(s)$,  $B(s)$ and $\alpha(s)$ are found from fitting of experimental data on elastic 
$pn\rightarrow pn$ scattering. 
For the effective lab momentum range of up  to $1.3$ GeV/c   the SAID parameterization\cite{SAID} of $pn$ amplitudes
can be used. The situation is more uncertain for the half-off-shell part of the $f^{NN,off}$ amplitude.  
In present calculations we use the following parameterization:
\begin{equation}
f^{NN,off} = f^{NN,on} e^{B(m_{off}^2 -m^2)},
\label{fnn_off}
\end{equation}
where $m^2_{off} = (M_d-E^\prime_{r} + q_0 )^2 - (p^\prime_r+q)^2$.  Overall we expect that our calculation will not 
be reliable in situations in which the contribution from the off-shell part of the rescattering is dominant.  
However in high $Q^2$ limit this contribution is only a small correction.

Completing this section it is worth to notice that in addition to the appearance of the $\Delta$ factor~(Eq.(\ref{Delta})) 
in GEA  which does not enter in conventional Glauber approximation (see detailed discussion in Ref.\cite{MS01}),  the new factor, 
$\frac{\sqrt{s(s-4m^2)}}  {2E^\prime_r |q|}$ entering the elastic FSI amplitude~(Eq.(\ref{a1_fsi})) is also unique to GEA.
Within conventional Glauber approximation, in which Fermi motion of the scatterers is neglected this factor is equal to 
one.  However within GEA it appears as a consequence of the  covariant form of the initial scattering amplitude.  
Calculation of this factors for our kinematics yields:
\begin{equation}
\frac {\sqrt{s(s-4m^2)}} {2E^\prime_r |q|} =  \frac {\sqrt{ ( {2-x\over x}Q^2 - m_D^2) ({2-x \over x}Q^2) }}  {2E^\prime_r |q|}
\end{equation}
which decreases with $x\to 2$.  Thus for the  $x>1$ and large $Q^2$ kinematics GEA predicts an additional suppression of FSI
as compared to the conventional Glauber approximation.

\subsubsection{Charge-Exchange Final State Interaction Amplitude}
To complete the calculation of the total amplitude we need to include the contribution from charge-exchange 
rescattering,  which can be obtained from Eq.(\ref{a1_fsi}) after the substitutions corresponding to Fig.1c.  
Namely, one needs to switch the  proton and neutron lines in the initial and intermediate states of the scattering, 
replace proton electromagnetic current by the neutron and $f_{NN}$ by the charge-exchange scattering 
amplitude $f_{NN}^{chex}$. One obtains:
\begin{eqnarray}
& & \langle s_f,s_r \mid A_{1,chex}^\mu\mid s_d\rangle  =     {i \sqrt{2}(2\pi)^{3\over 2}\over 4} \sum\limits_{s^\prime_f,s^\prime_r,s_i} 
\int {d^2p_r^\prime \over  (2\pi)^2}  \frac{\sqrt{2\tilde E^\prime_r}\sqrt{s(s-4m^2)}} {2\tilde E^\prime_r |q|}  \times \nonumber \\
& &  \ \ \ \ \ \  \langle p_f,s_f;p_r,s_r\mid f^{chex,on}(t,s)\mid \tilde p^\prime_r,s^\prime_r; \tilde p^\prime_f,s^\prime_f\rangle \cdot  
J_{n}^\mu(s^\prime_f,p^\prime_f;s_i,\tilde p^\prime_i) \cdot  \Psi_d^{s_d}(s_i,\tilde p^\prime_i,s^\prime_r,\tilde p^\prime_r) \nonumber  \\
& & -  {\sqrt{2}(2\pi)^{3\over 2}\over 2} \sum\limits_{s^\prime_f,s^\prime_r,s_1} {\cal P}\int {dp^\prime_{r,z}\over 2\pi} 
\int {d^2p_r^\prime \over  (2\pi)^2}  \frac{\sqrt{2E^\prime_r}\sqrt{s(s-4m^2)}} {2E^\prime_r |{\bf q}|} 
\times \nonumber \\
 & & \ \ \ \ \ \  {\langle p_f,s_f;p_r,s_r\mid f^{chex,off}(t,s)\mid p^\prime_r,s^\prime_r;p^\prime_f,s^\prime_f\rangle
\over  p^\prime_{r,z}- \tilde p^\prime_{r,z} }  J_{n}^\mu(s^\prime_f,p^\prime_f;s_i,p^\prime_i)
\cdot  \Psi_d^{s_d}(s_i,p^\prime_i,s^\prime_r,p^\prime_r).
\label{a1_chex}
\end{eqnarray}
Here the charge-exchange rescattering amplitude  $f_{NN}^{chex}$,  similar to the elastic FSI case  is taken from the experimental 
measurements.  The off-shell extrapolation of the rescattering amplitude is also done similar to Eq.(\ref{fnn_off}).
For numerical estimates we use the parameterization of Ref.\cite{GL}.

\subsubsection{Deuteron Wave Function}

\begin{figure}[th]
 \centering\includegraphics[scale=0.9]{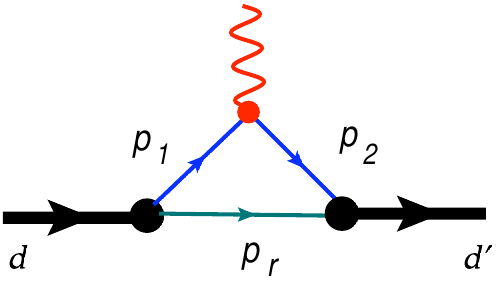}
  \caption{Elastic $\gamma^*d \rightarrow d^\prime$ diagram.}
\label{fig2}
\end{figure}

The deuteron wave function in Eq.(\ref{wf}) in general represents a solution of the  Bethe-Salpeter 
type equation.  To fix the normalization of the wave function  we need to relate an expression that 
contains the deuteron wave function (as it is defined in Eq.(\ref{wf})) to a  well defined quantity characterizing 
the deuteron. One such quantity is the deuteron elastic charge form-factor $G_C$, which at $Q^2\rightarrow 0$ 
limit approaches to one, i.e. $G_C(Q^2=0) = 1$ (see e.g. Ref.\cite{GG}).  
The latter could be related to the deutron elastic scattering amplitude as follows:
\begin{equation}
{1\over 4M_d} \sum\limits_{s^\prime_d = s_d=-1}^1 \langle p^\prime_d,s^\prime_d\mid A^{\mu=0}(Q^2)\mid  p_d, s_d\rangle
\mid_{Q^2\rightarrow 0} = G_{C}(0) = 1,
\label{charge}
\end{equation}
where $\langle p^\prime_d,s^\prime_d\mid A^\mu\mid  p_d, s_d\rangle$ is the elastic $\gamma^*d\rightarrow d^\prime$ 
scattering amplitude corresponding to the diagram of Fig.\ref{fig2}.

Applying the same effective Feynman diagram rules used above for  
$\langle p^\prime_d,s^\prime_d\mid A^\mu\mid  p_d, s_d\rangle$ one obtains:
\begin{eqnarray}
\langle p^\prime_d,s^\prime_d\mid A^\mu\mid  p_d, s_d\rangle = 
-\sum\limits_{p,n}\int {d^4 p_r\over i (2\pi)^4} \chi^{s^\prime_d,\dagger} \Gamma_{DNN}^\dagger
{\sh p_2 + m\over p_2^2 - m^2 +i\epsilon} 
\Gamma^\mu_{\gamma^* N} {\sh p_1 + m\over p_1^2 - m^2 +i\epsilon}\Gamma_{DNN}\chi^{s_d}\nonumber \\ 
\times {\sh p_r + m\over p_r^2 - m^2 +i\epsilon}. \ \ \ \ 
\label{el1}
\end{eqnarray}

Further derivations within the virtual nucleon approximation follow the similar to Secs.II.C.1 and II.C.2 steps. 
We first evaluate $dp_r^0$ integral   by the pole value of the spectator nucleon propagator, then separate 
the on- and off-shell parts in the numerator of  interacting nucleon propagators and introduce 
deuteron wave function according to Eq.(\ref{wf}). In this case the electromagnetic current of the nucleon 
is  fully off-shell since the struck nucleon is bound in both initial and final states of the reaction.
This results in the following expression for the elastic scattering amplitude:
\begin{eqnarray}
\langle p^\prime_d,s^\prime_d\mid A^\mu\mid  p_d, s_d\rangle = 4\sum\limits_{p,n}\sum\limits_{s_2,s_1,s_r}
\int d^3 p_r \Psi_d^{s^\prime_d\dagger}(s_2,p_2,s_r,p_r)
\bar u(p_2,s_2)\left[I + {E_2^{off}-E_2^{on}\over 2m}\gamma_0\right]\nonumber \\
\Gamma^\mu_{\gamma^*N}\left[I + {E_1^{off}-E_1^{on}\over 2m}\gamma_0\right]u(p_1,s_1) 
\Psi_d^{s_d}(s_1,p_1,s_r,p_r). \ \ \ \ 
\label{el2}
\end{eqnarray}
Neglecting the second order off-shell terms in the above equation (i.e. $({E^{off}-E^{on}\over 2m})^2$) one 
obtains:
\begin{equation}
\langle p^\prime_d,s^\prime_d\mid A^\mu\mid  p_d, s_d\rangle = 4\sum\limits_{p,n}\sum\limits_{s_2,s_1,s_r}
\int d^3 p_r \Psi_d^{s^\prime_d\dagger}(s_2,p_2,s_r,p_r) \tilde J^\mu_{N} 
 \Psi_d^{s_d}(s_1,p_1,s_r,p_r),
\label{el3}
\end{equation}
where
\begin{equation}
 \tilde J_{N}^\mu(s_2,p_2;s_1,p_1) = J_{N,on}^\mu(s_2,p_2;s_1,p_1)+J_{N,off}^\mu(s_2,p^{off}_2;s_1,p_1)
+ J_{N,off}^\mu(s_2,p_2;s_1,p^{off}_1).
\end{equation}
Here the on- and off- shell parts of electromagnetic current are defined in Eq.(\ref{J_on}) and (\ref{J_off}).
In the above equation the notation $p^{off}$ in the argument of 
$J_{N,off}$ indicates which nucleon is considered as off-shell.

Using now the fact that for the proton and neutron $F_{1p(n)}(Q^2=0)=1(0)$ and 
using Eqs.(\ref{J_on},\ref{J_off}) one obtains:
\begin{eqnarray}
\tilde J^{\mu=0}_{p}\mid_{Q^2\rightarrow 0} & = & 2 E_1^{off}\nonumber \\ 
\tilde J^{\mu=0}_{p}\mid_{Q^2\rightarrow 0} & = & 0.
\end{eqnarray} 
Using these relations and inserting Eq.(\ref{el3}) into Eq.(\ref{charge}) one obtains
\begin{equation}
\sum\limits_{s_d=-1}^{1}\int \mid \Psi^{s_d}_d(p)\mid^2 {2E_{off}\over M_d}d^3p = 1
\label{norm}
\end{equation}
where $E^{off} = M_d - \sqrt{m^2+p^2}$. It is worth mentioning that above normalization coincides 
with the normalization obtained from the baryon number conservation sum rule\cite{BN,LP,SFS,CL,
gdpn,SSS02,polext} for deep inelastic scattering off the deuteron:
\begin{equation}
\int  |\Psi_d(\alpha,p_t)|^2 \alpha d^3p = 1
\end{equation}
where $\alpha = {M_d-\sqrt{m^2+p^2} - p_z\over m}$ is the light-cone momentum fraction of the deuteron carried by 
the struck nucleon. As it was mentioned in Sec.II.B in virtual nucleon approximation due to 
unaccounted non-nucleonic degrees of freedom the wave function defined according to Eq.(\ref{wf}) will 
not satisfy the energy-momentum sum rule which expresses the requirement 
that the sum of the light-cone momentum fractions of  all the constituents of the nucleus equals to one.

For numerical estimates we model the deuteron wave function to satisfy Eq.(\ref{norm}) in the following 
form\cite{SFS,polext}:
\begin{equation}
\Psi_d(p) = \Psi^{NR}_d(p) {M_d\over 2(M_d - \sqrt{m^2+p^2})} 
\label{wfmod}
\end{equation}
which provides a smooth transition to the nonrelativistic wave function $\Psi^{NR}$ in the small momentum limit.

\subsubsection{Total amplitude and the differential cross section}
The total scattering amplitude consists of the sum of PWIA, forward and charge-exchange FSI amplitudes:
\begin{equation}
 \langle s_f,s_r \mid A^\mu\mid s_d\rangle  = 
 \langle s_f,s_r \mid A_{0}^\mu\mid s_d\rangle +  \langle s_f,s_r \mid A_{1}^\mu\mid s_d\rangle +  
\langle s_f,s_r \mid A_{1,chex}^\mu\mid s_d\rangle . 
\label{A_tot}
\end{equation}
Using this  amplitude the differential cross section is calculated as follows:
\begin{equation}
\frac {d\sigma}{dE_e^\prime, d\Omega_{e^\prime} d p_f d\Omega_f} = \frac { \alpha^2E^\prime_e}{q^4 E_e} 
\cdot {1\over 6} \sum\limits_{s_f,s_r,s_d,s_1,s_2} \frac{\mid J_e^\mu J_{d,\mu}\mid^2}{2M_d E_f}{p_f^2\over 
\mid {p_f\over E_f} + {p_f - q cos(\theta_{p_f,q})\over E_r}\mid}
\label{crs}
\end{equation}
where 
\begin{equation}
J_e^\mu = \bar u(k_2,s_2)\gamma^\mu u(k_1,s_1)
\label{Je}
\end{equation}
and
\begin{equation}
J_d^\mu = { \langle s_f,s_r \mid A^\mu\mid s_d\rangle\over \sqrt{2(2\pi)^3 2 E_r}}.
\label{Jd}
\end{equation}

For numerical estimates we use the electromagnetic current of the nucleon in the 
form
\begin{equation}
\Gamma^\mu = F_1(Q^2)\gamma^\mu + {F_2(Q^2) \over 2m}i \sigma^{\mu,\nu}q_\nu,
\label{Gamma_N}
\end{equation}
where $F_1$ and $F_2$ are Dirac form-factors and for their evaluation the available phenomenological 
parameterizations are used\cite{Kelly}.
For the deuteron wave function we use the approximation of Eq.(\ref{wfmod}) with the non-relativistic 
wave function calculated based on  the Paris potential\cite{Paris}. The $pn$ scattering amplitude is 
parameterized in the form of Eq.(\ref{fnn_on}) and its off-shell extrapolation in the form of 
Eq.(\ref{fnn_off}). Also, for $f_{pn}$ in the  lower momentum range ($p_{lab}\le 1.3$~GeV/c) we use 
the SAID parameterization\cite{SAID} based on the $pn$ scattering phase-shifts.

\section{Observables}
The main quantity  which we will consider for numerical estimates is the 
ratio of  the calculated cross section which includes total amplitude of Eq.(\ref{A_tot})  to the 
cross section calculated within PWIA:
\begin{equation}
R = {\sigma\over \sigma^{PWIA}}
\label{R}
\end{equation}
where $\sigma \equiv = {d\sigma\over dE_e^\prime, d\Omega_{e^\prime} d p_f d\Omega_f}$.
This ratio allows us to clearly distinguish  between kinematics in which PWIA dominates $R\approx 1$  from 
kinematics in which  FSI is dominated by  screening  $R < 1$  or single rescattering $R>1$ effects 
(see e.g. \cite{edepn95,dshadow,gea}).

Considering the numerical estimates of the ratio $R$ we will discuss  four main effects 
which  characterize  our present theoretical approach. These are
the unfactorization of the electromagnetic interaction in the FSI amplitude, the   off-shell effects in 
the final state interaction, the effects of charge-exchange rescatterings  and the off-shell effects in 
the electromagnetic interaction of the  bound nucleon.

In our estimates we will  study the dependence of $R$
on the angle of the recoil neutron relative to $\vec q$ for different values of  neutron momenta.
We will perform our calculations for two values of $Q^2$~ (2~GeV$^2$ and  6~GeV$^2$) which will allow us 
to also assess  the $Q^2$ dependence of the considered effects.

Finally,  we will present the comparisons  with the first experimental data on the deuteron electrodisintegration 
at large $Q^2$.

\begin{figure}[th]
 \centering\includegraphics[scale=0.5]{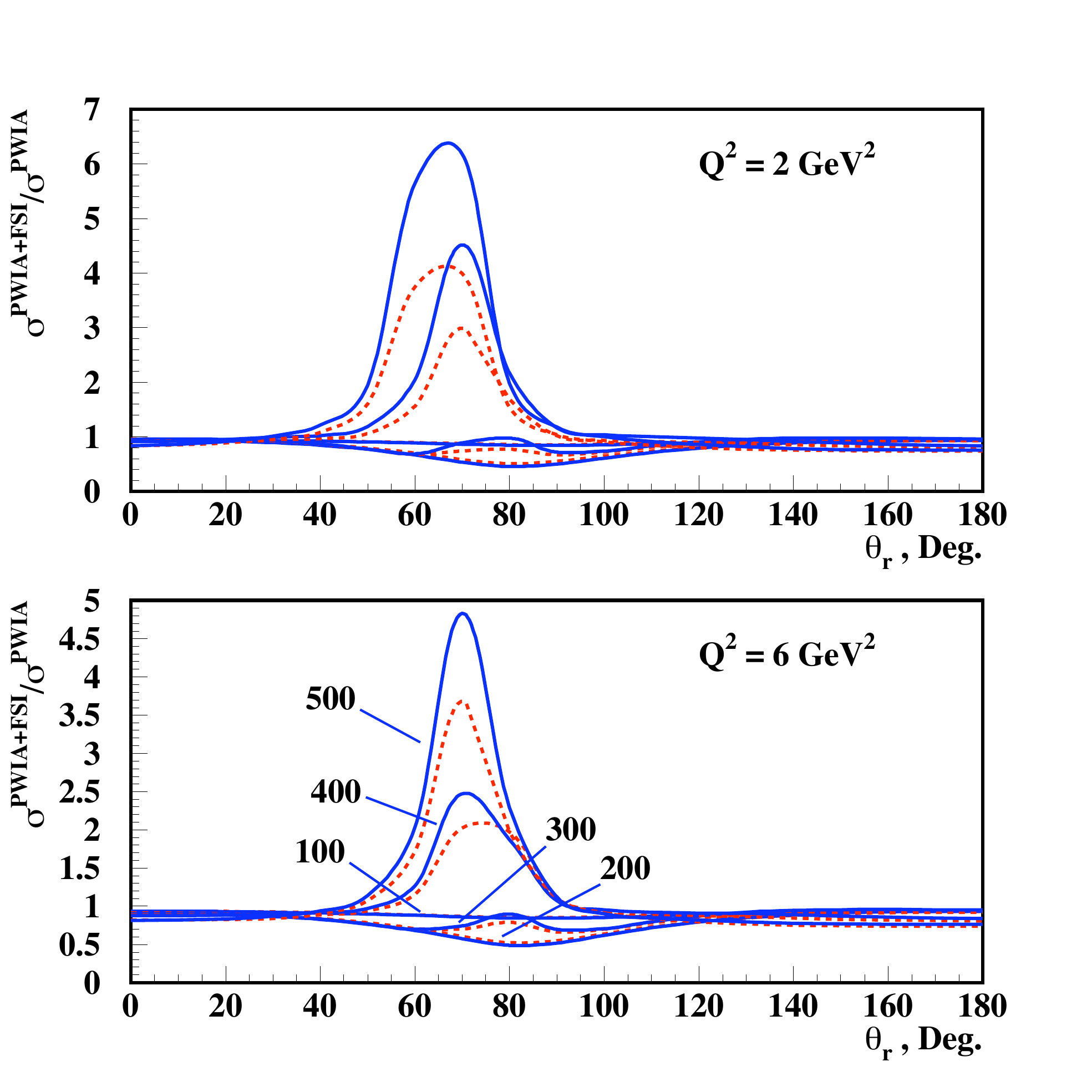}
  \caption{Dependence of ratio $R$ on the recoil angle of the neutron for different values of 
  $p_r=100,200,300,400,500$~MeV/c and $Q^2=2,6$~GeV$^2$. Solid line - unfactorized and dashed line 
  - factorized approximations.}
\label{fig3}
\end{figure}

\subsection{Nonfactorization effects}

In Fig.\ref{fig3} we compare the calculations of $R$ with and without factorization approximation for the electromagnetic 
current in the FSI amplitude.  The factorization approximation will result in the  so called distorted wave impulse 
approximation~(DWIA) widely  used in the literature.

As the figure shows the factorization (or DWIA) approximation is applicable  for up to $p_r \le 300$~MeV/c or for the 
kinematics in which the FSI amplitude  is smaller than the PWIA  term.  The factorization approximation breaks down in 
kinematics dominated by the rescattering process at $p_r\ge 400$~MeV/c.  As the comparisons show,  the unfactorization 
predicts a  larger FSI amplitude which can be understood based on the fact that in this case the electromagnetic 
current which enters in the rescattering amplitude of Eq.(\ref{a1_fsi})  is defined at smaller values of bound nucleon momenta
than the electromagnetic current in the PWIA term (Eq.(\ref{A0_3})). 
 
Fig.\ref{fig3}  shows also that the factorization approximation is $Q^2$ dependent and somewhat improves with an 
 increase of $Q^2$.  This is a  rather  important feature which should be taken into account 
in color transparency studies~(CT) for double scattering kinematics, in which case the $Q^2$ dependence of the 
peak of the ratio $R$ is studied in order to extract the CT signal (see e.g \cite{edepn95,dshadow,drsc93}). 
Our comparisons in  Fig.\ref{fig3} shows  that the unfactorized approximation should be used as a baseline 
for identification of the CT signature in the $Q^2$ dependence of the FSI contribution of  the 
deuteron break-up cross section.

\begin{figure}[th]
  \centering\includegraphics[scale=0.5]{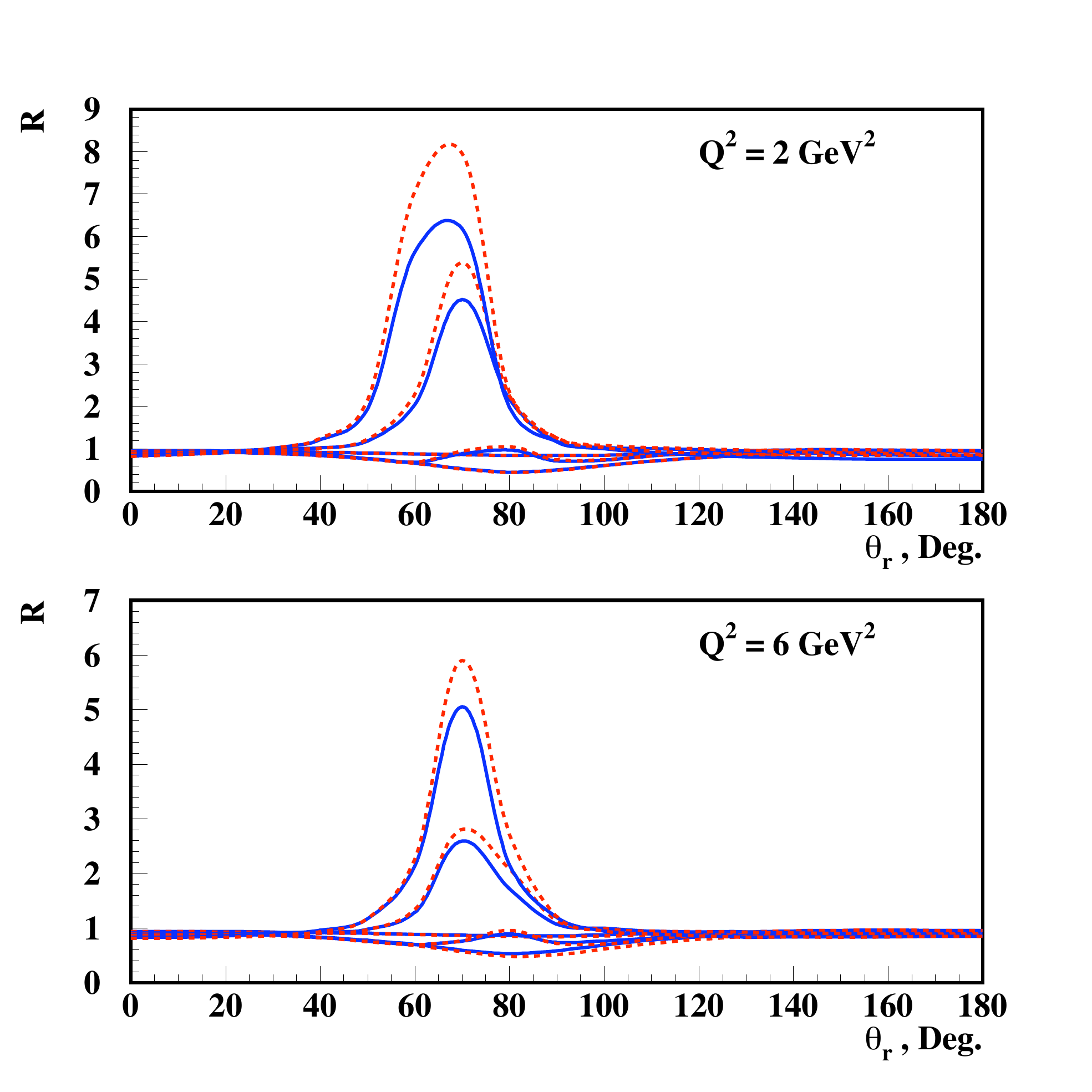}
  \caption{Dependence of ratio $R$ on the recoil angle of the neutron for different values of 
  $p_r$ and $Q^2=2,6$~GeV$^2$. The recoil neutron momenta are the same as in Fig.3. 
Solid line - unfactorized  calculation with the pole term only in the FSI amplitude,  dashed line 
- unfactorized approximations with both pole  and principal value terms in the FSI amplitude.}
\label{fig4}
\end{figure}

\begin{figure}[th]
  \centering\includegraphics[scale=0.5]{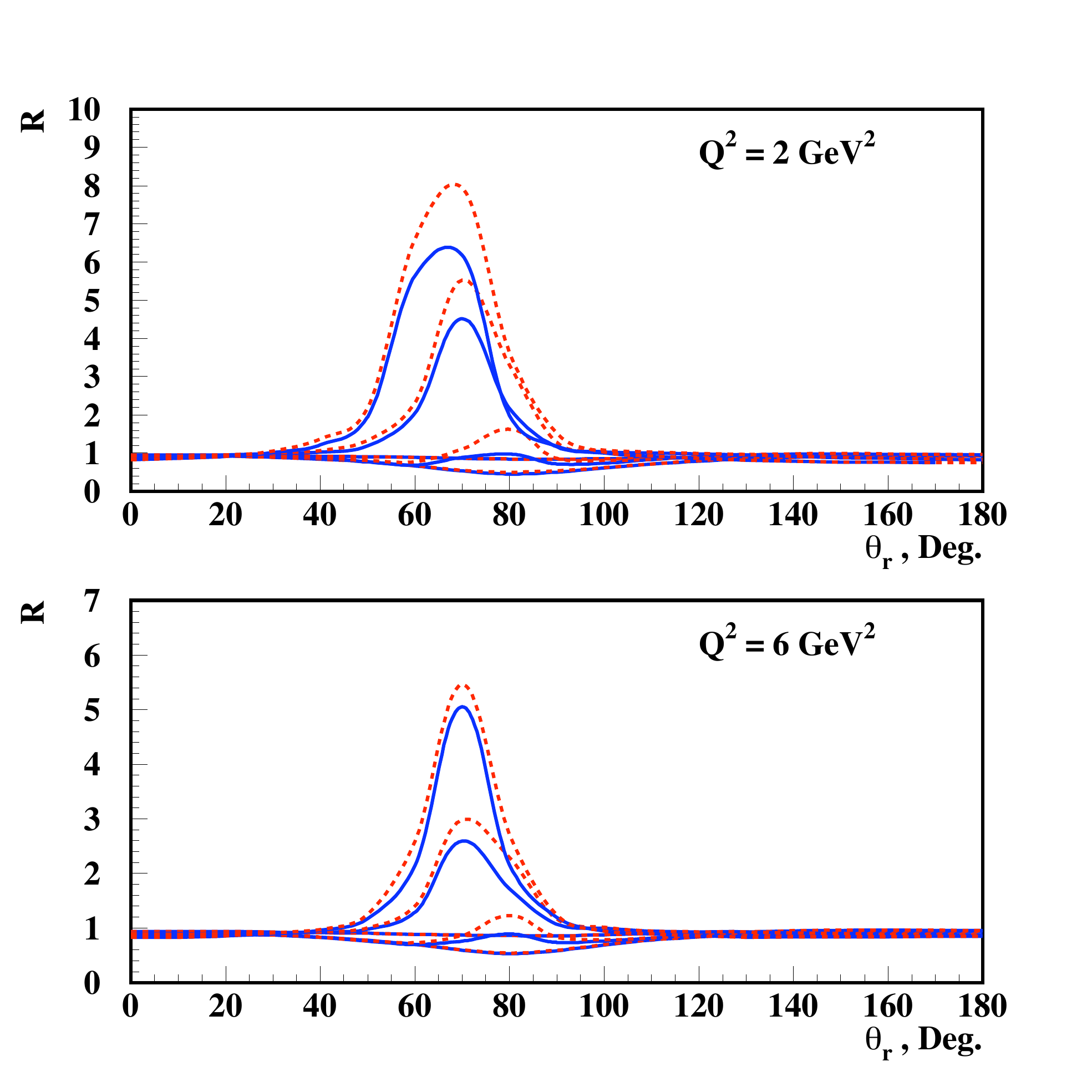}
  \caption{Dependence of ratio $R$ on the recoil angle of the neutron for different values of 
  $p_r$ and $Q^2=2,6$~GeV$^2$. The recoil neutron momenta are the same as in Fig.3. 
Solid line - unfactorized  approximation with the pole term only  forward $pn\rightarrow pn$  rescattering,  
dashed line -  unfactorized approximation including the pole  terms for both 
forward $pn\rightarrow pn$  and  charge-exchange $pn\rightarrow np$ rescattering amplitudes.}
\label{fig5}
\end{figure}

\subsection{Off-Shell FSI Effects}
Next we consider the  contribution to the FSI amplitude 
due to  the principal value part  of Eq.(\ref{a1_fsi}).  
This term depends on the half-off shell $NN$ scattering amplitude which is a  
largely unknown quantity. Therefore the reliability of our  calculations depends on the  
magnitude of the principal value  term. 
In Fig.\ref{fig4} we compare the calculations in which only the pole (on-shell) term of the FSI amplitude  
is included with the calculations which contain both pole (on-shell) and principal value (off-shell) terms of the 
FSI amplitude. For the half-off shell $f_{pn}$ amplitude we use the approximation of Eq.(\ref{fnn_off}).
An important observation can be  made from Fig.\ref{fig4} is that the off-shell rescattering effects diminish with 
an increase of $Q^2$. This is consistent with our earlier observation\cite{Day,revsrc} that the distances 
that the virtual nucleon propagates before the rescattering decreases significantly  with an increase of 
$Q^2$ at fixed values of $x$.

\begin{figure}[th]
  \centering\includegraphics[scale=0.5]{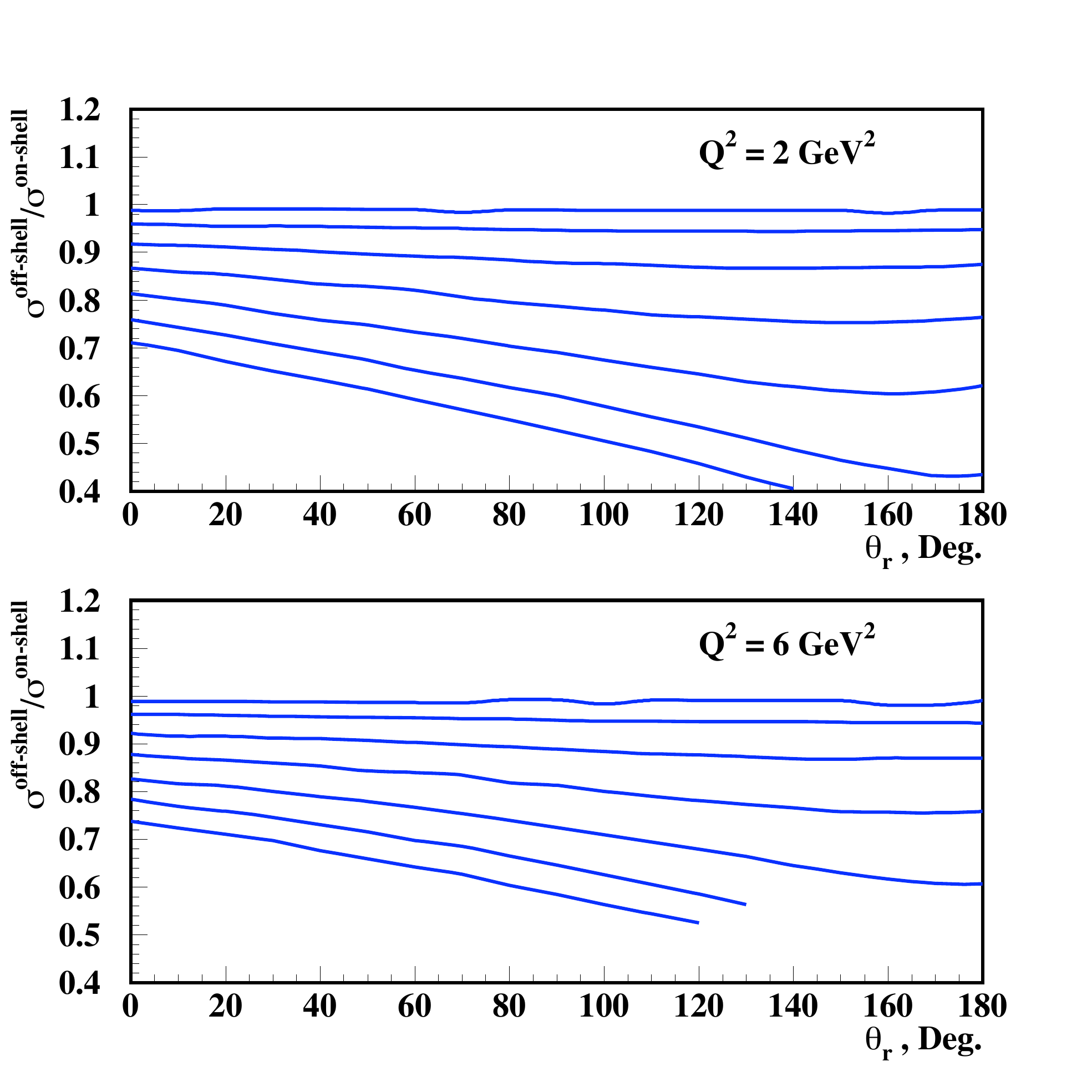}
  \caption{Ratio of PWIA cross sections calculated using on-shell and off-shell approximations for 
the electromagnetic current of the  proton. The curves from top to bottom correspond to 
recoil neutron momenta 100, 200, 300, 400, 500, 600 and 700~MeV/c respectively.}
\label{fig6}
\end{figure}

\subsection{Charge Exchange Rescattering Effects}

Due to the fact that the charge-exchange rescattering amplitude is predominantly real it will interfere mainly with the 
real part of the forward elastic FSI amplitude which is a small parameter at large energies.  
One can see from  Fig.\ref{fig5} that the  charge-exchange rescattering dominates at kinematics in which 
the rescattering defines the overall magnitude of the cross section. 

However it is a rather well known fact that, due to the dominant pion-exchange nature of charge-exchange $pn$ scattering, 
its cross section decreases linearly with an increase of  the invariant energy $s$ as compared to the 
forward $pn$ elastic scattering cross section.
As a result one expects that with an increase of $Q^2$ the charge exchange rescattering term will become a small 
correction.  This can be seen from the calculation for $Q^2=6$~GeV$^2$ kinematics in Fig.\ref{fig5}.

\subsection{Off-Shell Electromagnetic  Interaction Effects}

The above evaluations of the $d(e,e'p)n$ cross sections demonstrate that in the large $Q^2$ limit the property of 
the scattering is defined mainly by the PWIA  and forward angle on-shell FSI terms.  
The angular distribution has a very characteristic shape in which one can identify kinematics dominated by 
PWIA or FSI.
Calculations also show that the forward or backward kinematics of the recoil nucleon are best suited 
for isolating PWIA term which subsequently allows us to study the structure of the electromagnetic current 
of the bound nucleon and the 
deuteron wave function at large values of internal momenta.

We now concentrate on the effects related to the fact that the proton in the deuteron becomes increasingly 
off-shell at large values of recoil neutron momenta in forward/backward directions.

As it follows from Eqs.(\ref{J_on},\ref{J_off}) and (\ref{J_N}) the off-shell part of the electromagnetic current will diminish 
the overall magnitude of the electromagnetic interaction.  Since the off-shellness  grows with an increase of 
the initial momentum of the struck nucleon it will result in the suppression of the electromagnetic interaction strength
of  deeply bound nucleons.
As it follows from Fig.\ref{fig6} the off-shell effects have a weak $Q^2$ dependences and  to disentangle them from the 
effects related to the high momentum component of the deuteron wave function would require measurements covering 
an extended angular and recoil momentum range. Fig.\ref{fig6} also shows that the forward direction of 
recoil nucleon momenta represents the most optimal kinematic region 
for  minimizing the off-shell effects  for electromagnetic interaction.

Note that polarization measurements will provide additional observables for 
separating current and wave function effects. For example, measurements of the cross section asymmetries similar to 
Ref.\cite{Stefen} at large recoil momenta will be more sensitive to the structure of the electromagnetic current 
since wave function effects largely cancel out in the ratios defining the asymmetry.

\subsection{Comparison with Experimental Data}

In the last few years  three experiments\cite{Ulmer,Kim3,Werner} produced first data
at relatively large $Q^2$ kinematics.
 
\begin{figure}[th]
  \centering\includegraphics[scale=0.5]{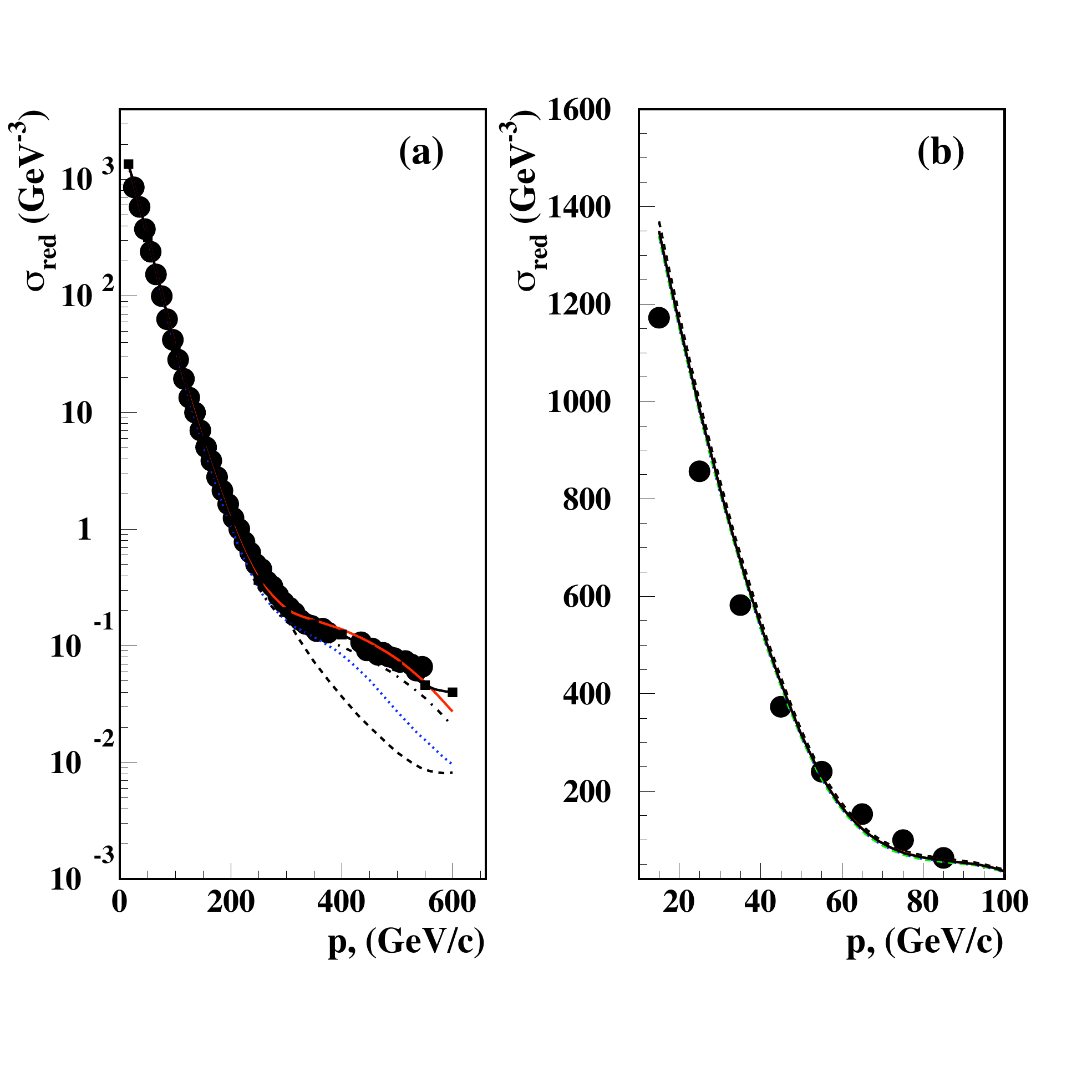}
  \caption{Missing momentum dependence of the reduced cross section. The data are from Ref.\cite{Ulmer}.
Dashed line - PWIA calculation, dotted line - PWIA+ only pole term of forward FSI,  dash-dotted line  -  PWIA+ forward FSI,  
solid line - PWIA + forward and charge exchange FSI, and solid line with squares - same as the previous solid 
line, added the  contribution from the mechanism in which the proton is a spectator and the neutron was struck 
by the virtual photon.}
\label{fig7}
\end{figure}

The first experiment\cite{Ulmer} probed the $Q^2 = 0. 665$~GeV$^2$ and $x\approx 1$ kinematics and provided 
very accurate data. The  measured value of $Q^2$ is marginal
for the application of GEA. However as Fig.\ref{fig7} shows the comparison 
yields a  surprisingly good agreement with the data.   Fig.\ref{fig7} compares the reduced cross section defined as follows:
\begin{equation}
\sigma_{red} = \frac {d\sigma}{dE_e^\prime, d\Omega_{e^\prime} d p_f d\Omega_f} \cdot 
{\mid {p_f\over E_f} + {p_f - q cos(\theta_{p_f,q})\over E_r}\mid\over \sigma_{CC1}\cdot p_f^2}
\label{red}
\end{equation}
where the differential cross section is defined according to Eq.(\ref{crs}) and $\sigma_{CC1}$ is the off-shell 
electron-proton  cross section defined in Ref.\cite{DeForest}.
Note that in this calculation for $f_{pn}$ amplitude we use 
SAID\cite{SAID} parameterization for both elastic and charge-exchange $pn$ scatterings
which fit the elastic $pn$ scattering data for lab momenta up to 1.3 GeV/c.  Because of the relatively low 
energy and momentum transfers involved in the reaction the calculation shows a substantial contribution 
from the off-shell as well as charge exchange  parts of the FSI amplitude (difference between dotted, 
dash-dotted  and solid lines in Fig.\ref{fig7}).

It is worth noting that in agreement with the observations of Refs.\cite{Werner,Sabine4}  the calculation overestimates
the low recoil momentum part of the cross section (Fig.\ref{fig7}b), where we expect that theoretical uncertainties 
are well under control. Our preliminary estimates demonstrate that this discrepancy could be accounted for 
by inclusion of the contribution of two-photon exchange effects in  the overall amplitude of the scattering\cite{Misakpro2}.

The second experiment\cite{Kim3}) covered the  $Q^2$ range from $2-5$~GeV$^2$. However, due to the  low 
statistics the data are integrated over the ranges of the recoil nucleon momenta.

\begin{figure}[th]
  \centering\includegraphics[scale=0.5]{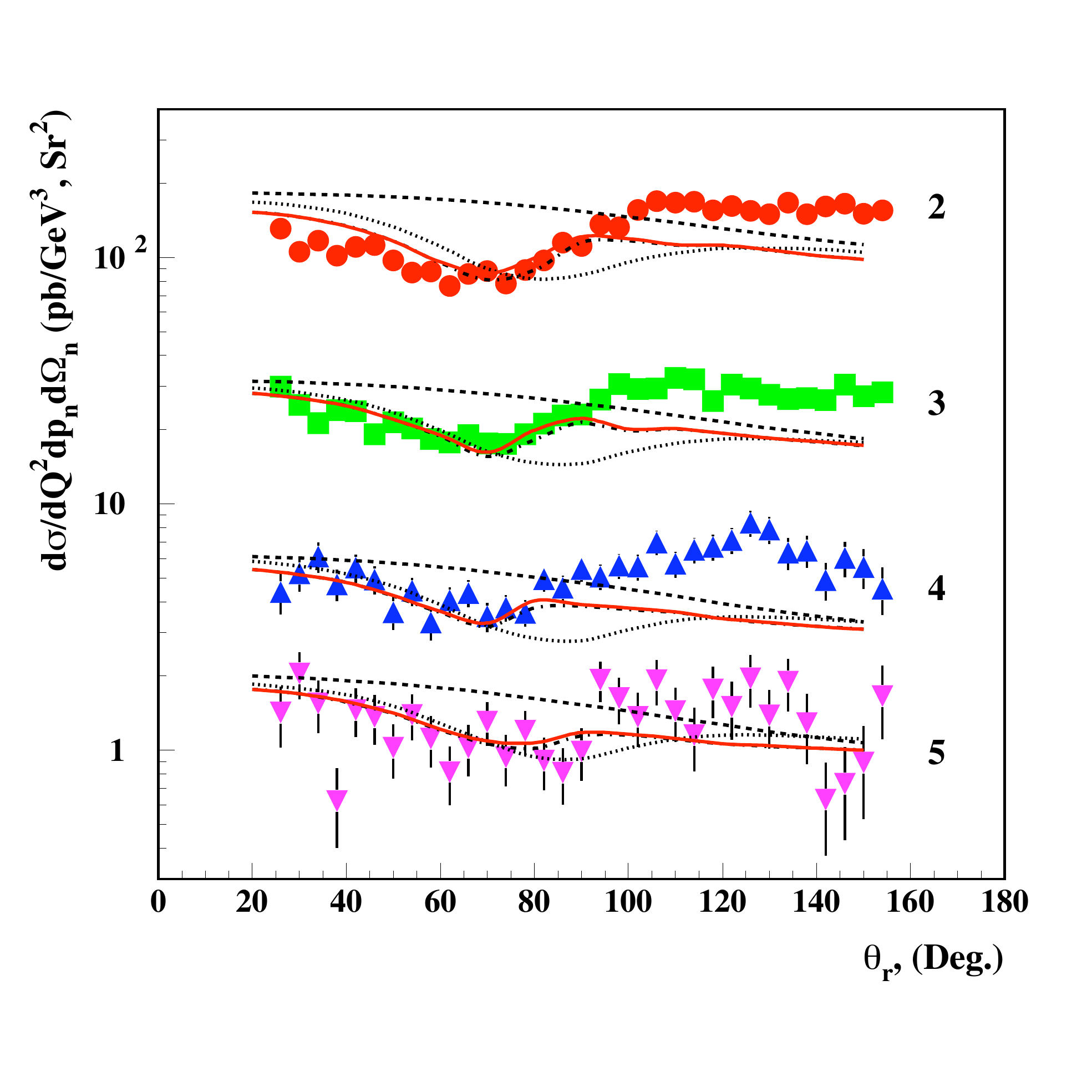}
  \caption{ Dependence of the differential  cross section on the direction of the recoil neutron momentum. 
The data are from Ref.\cite{Kim3}.
Dashed line - PWIA calculation, dotted line - PWIA+ pole term of forward FSI,  dash-dotted line  -  PWIA+forward FSI,  
solid line - PWIA + forward and charge exchange FSI.  The momentum of the recoil neutron is 
restricted to $200 < p_r < 300$~MeV/c.   The labels 2, 3, 4 and 5 correspond to the following values of 
$Q^2 = 2\pm 0.25;  3\pm 0.5; 4\pm 0.5; 5\pm 0.5$~GeV$^2$.}
\label{fig8}
\end{figure}

\begin{figure}[th]
  \centering\includegraphics[scale=0.5]{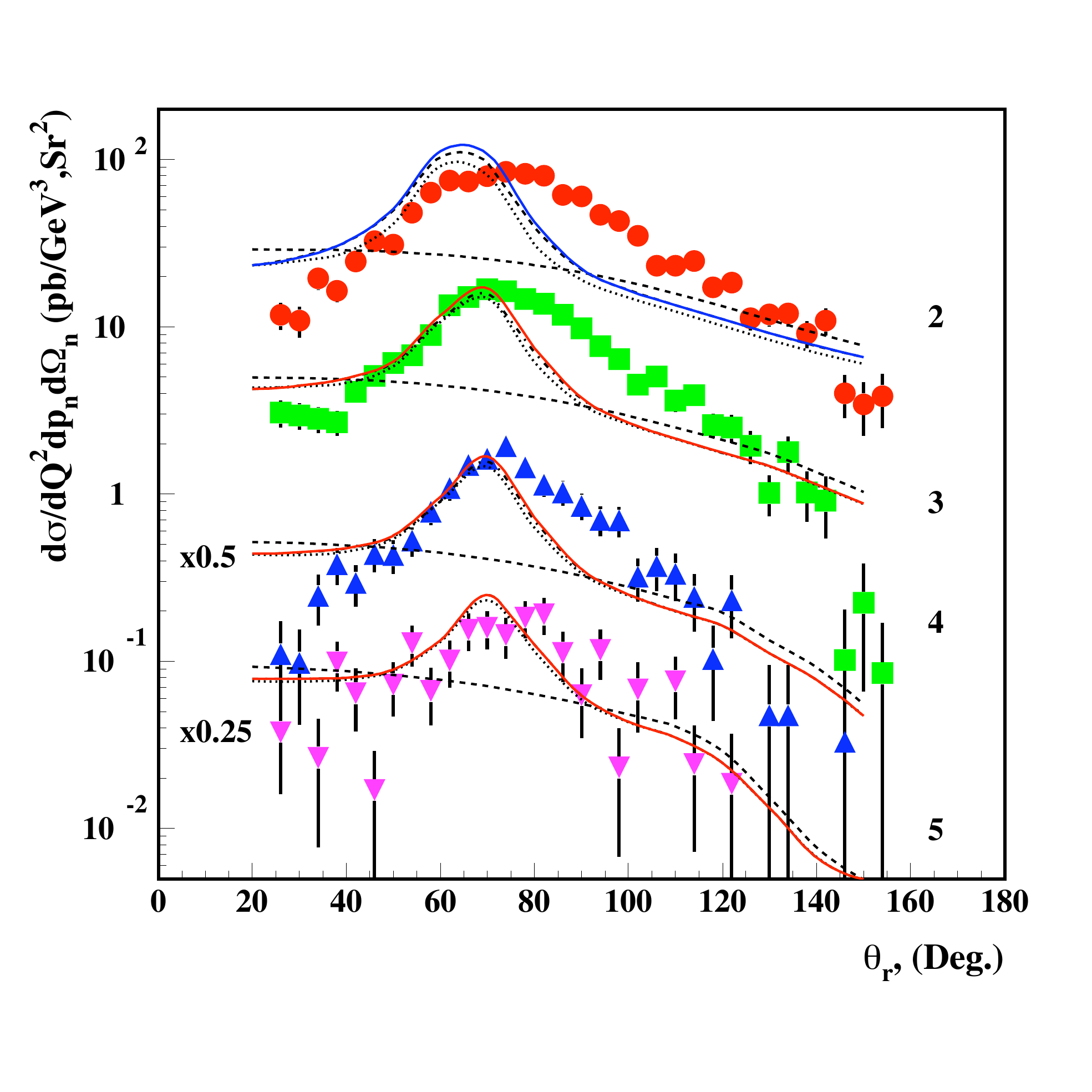}
  \caption{ Dependence of the differential  cross section on the direction of the recoil neutron momentum. 
The data are from Ref.\cite{Kim3}.
Dashed line - PWIA calculation, dotted line - PWIA+ pole term of forward FSI,  dash-dotted line  -  PWIA+forward FSI,  
solid line - PWIA + forward and charge exchange FSI.  The momentum of the recoil neutron is 
restricted to $400 < p_r < 600$~MeV/c.   The labels 2, 3, 4 and 5 correspond to the following values of 
$Q^2 = 2\pm 0.25;  3\pm 0.5; 4\pm 0.5; 5\pm 0.5$~GeV$^2$.  The data sets and calculations for 
``4'' and ``5'' are multiplied by  0.5 and 0.25 respectively.}
\label{fig9}
\end{figure}

Comparing with these data we performed the same kinematic integrations as the experiment did.  
The results are shown in Fig.\ref{fig8} and \ref{fig9}. Despite these integrations we still 
can make  several important observations from these comparisons.

\begin{itemize}
\item First, the angular distribution clearly exhibits   an eikonal feature, with the minimum (Fig.\ref{fig8}) 
or maximum (Fig.\ref{fig9})  at transverse kinematics 
due to the  final state interaction.  The most important result is that the maximum of FSI is at
recoil angles of $70^0$   in agreement with the GEA prediction of Ref.\cite{gea}. Note that the  conventional 
Glauber theory predicted $90^0$ for the FSI maximum.

\item The disagreement of the calculation with the data at  $\theta_r > 70^0$ 
appears to be due to the isobar contribution at the 
intermediate state of the reaction. This region corresponds to $x<1$ and it is   
kinematically  closer to the threshold  of $\Delta$-isobar 
electroproduction. The comparisons also indicate 
that the relative strength of the  $\Delta$-isobar contribution diminishes with an increase of 
$Q^2$ and at neutron production angles $\theta_r\rightarrow 180^0$.  

\item The forward direction of the recoil nucleon momentum, being far from the $\Delta$-isobar threshold, 
exhibits a relatively small contribution due to FSI.  This indicates that the forward recoil angle 
region is best suited for studies of  PWIA properties of the reaction such as the off-shell 
electromagnetic current and deuteron wave function. 
\end{itemize}

Finally, the results of the experiment of Ref.\cite{Werner} are currently in the final stages of analysis.
Since in this experiment the disintegration reaction is measured at forward recoil angles and at $Q^2$ up to 3.5~GeV$^2$, we 
expect that it will allow us to further check the validity of our claim that the 
forward angular  region is  best suited for studies of the  PWIA properties of the reaction.

\section{Conclusions}

Within the virtual nucleon approximation we developed a theoretical framework for calculation of 
high $Q^2$ exclusive 
electrodisintegration of the deuteron at large values of recoil nucleon momenta.
The scattering amplitude is derived based on generalized eikonal approximation, in which case each 
amplitude is calculated based on effective Feynman diagram rules. Because of the covariant formulation of the 
problem the electromagnetic current is gauge invariant from the beginning. By isolating the off-shell part 
in the electromagnetic current we introduced an approach which allows us to express the bound nucleon current 
separately through the on-shell and off shell parts.

Next, we derived the final state interaction amplitude based on GEA and expressed it through the on-shell 
and off-shell 
rescattering parts.  No factorization approximation is assumed in the calculation of the FSI amplitude.
The calculation of FSI also includes the amplitude due to 
the charge-exchange final state interaction.

We performed numerical analysis of the obtained formulae to identify the level of uncertainties due to  
the factors included in the calculations. Overall, our conclusion is that with an increase of $Q^2$ all the uncertainties 
related to the off-shell FSI and charge exchange rescattering became small and the total scattering amplitude 
is determined by the PWIA and on-shell elastic $NN$ rescattering.

We compared our calculations with the first experimental data at large $Q^2$  deuteron electrodisintegration.
These comparisons revealed a rather surprising agreement with low $Q^2=0.665$~GeV$^2$ data  which indicates the wider 
range of applicability of the present approximation.

Comparisons with higher $Q^2$ data ($\ge 2$~GeV$^2$)  at the wider range of recoil nucleon momenta 
and angles demonstrate the important role that the intermediate $\Delta$-Isobar production plays in 
electrodisintegration reaction at backward angles close to the $\Delta$ production threshold.

However, at forward recoil angles  where FSI effects are restricted,  the calculations show 
greater sensitivity to  the PWIA structure of the electrodisintegration reaction.  Further experiments 
will allow us to confirm that this region is most suitable for probing the bound nucleon electromagnetic 
current and the deuteron wave function at small distances.

\acknowledgments
I'm  thankful  to Drs.~Werner Boeglin, Mark Strikman and 
late Kim Egiyan for numerous discussions on the physics of deuteron break-up.
I regret that this work was completed too late for Kim to see the importance of  
his deuteron break-up experiment. This work is supported by the  U.S. Department of Energy Grant 
under Contract DE-FG02-01ER41172.

\end{document}